\documentclass[twoside,12pt]{article}
\input{epsf.tex}
\usepackage{rotating}
\newcommand{\thishepex} {hep-ex/0401034}

\newcommand{\epemtohad} {$e^+e^-\rightarrow~{\mathrm{hadrons}}$}
\newcommand{\csepemtohad} {$\sigma(e^+e^-\rightarrow~{\mathrm{hadrons}})$}
\newcommand{\epem} {e^+e^-}
\newcommand{\qqbar} {q\bar{q}}

\newcommand{\alphaQED} {\alpha_{\mathrm{QED}}}
\newcommand{\LambdaQCD} {\Lambda_{\mathrm{QCD}}}
\newcommand{\dahad}    {\Delta\alpha^{{\mathrm{(had)}}}}
\newcommand{\gminustwo} {$(g-2)_\mu$}
\newcommand{\cspeak}   {\sigma^0_{\mathrm{had}}}
\newcommand{\ssqW}     {\sin^2\theta_W}
\newcommand{\csqW}     {\cos^2\theta_W}

\newcommand{\sigmaNP}  {\sigma^{\mathrm{NP}}}

\newcommand{\GeV}  {{\mathrm{GeV}}}
\newcommand{\pb}   {{\mathrm{pb}}}
\newcommand{\lep} {{\sc Lep}}
\newcommand{\lepone} {{\sc Lep~1}}
\newcommand{\leptwo} {{\sc Lep~2}}
\newcommand{\lepewwg} {{\sc LepEWWG}}
\newcommand{\zfitter} {{\sc Zfitter}}
\newcommand{\msmlib} {{\sc Msmlib}}

\newcommand{\ySM} {y^{\mathrm{SM}}}
\newcommand{\yNP} {y^{\mathrm{NP}}}
\newcommand{\alphabest} {\alpha_{\mathrm{best}}}
\newcommand{\alphanet} {\alpha_{\mathrm{net}}}

\newcommand{\Zprime}   {Z^\prime}
\newcommand{\MZprime}  {M_{Z^\prime}}
\newcommand{\sbone}  {\tilde{b}_1}
\newcommand{\Msbone} {M_{\tilde{b}_1}}
\newcommand{\gluino} {\tilde{g}}
\newcommand{\Mgluino} {M_{\tilde{g}}}
\newcommand{\cossbone} {\cos\theta_{\tilde{b}_1}}

\newcommand{\etal}  {{\em et~al.}}
\newcommand{\ie}    {{\em i.e.}}
\begin{document}
\rightline{\tt \thishepex}
\rightline{\tt nuhep-exp/04-01}
\begin{center}
{\Large Apparent Excess in \epemtohad}
\vskip 0.25in
Michael Schmitt
\vskip 0.125in
{\small
Northwestern University
}
\vskip 0.125in
January 22, 2004
\end{center}
\vskip 0.25in%
\begin{center}
{\bf Abstract}
\vskip 0.05in
\parbox[c]{5.5in}
{\small
We have studied measurements of the cross section for \epemtohad\
for center-of-mass energies in the range 20--209~GeV.
We find an apparent excess over the predictions of the Standard Model
across the whole range amounting to more than~$4\sigma$.  
As an example, we compare the data to predictions  for a light 
scalar down-type quark which fit the excess well.
}
\end{center}
%
\section{Introduction}
\par
Measurements of the inclusive cross section for \epemtohad\
have been a staple of $\epem$ collider experiments for many years.
They have been essential for testing the predictions of QCD, and
for measurements of the properties of the $Z$ boson which underpin
the electroweak sector of the Standard Model (SM).  It is often said
that the SM survives all confrontations with data, and indeed, this
agreement is a very important constraint on models for new physics
beyond the SM.  For example, one of the strengths of the 
Minimal Supersymmetric extension of the Standard Model (MSSM)
is the fact that it does not disturb this agreement though
its parameters may be varied over wide ranges.
\par
It has been noted several times that the \leptwo\ measurements
of \epemtohad\ tend to exceed the predictions of the SM~\cite{lep2exceed}.
It is also well known that the hadronic cross section at the $Z$~peak, 
$\cspeak$, slightly exceeds the SM fit~\cite{lepewwg,lepewwgZ}.   
Neither of these excesses is statistically significant.
\par
We have examined the data on \epemtohad\ from experiments that ran
below the $Z$ peak using a compilation amassed by Zenin \etal,~\cite{cscomp},
and find that these measurements, taken together, also are in excess
of the SM prediction.  We developed a likelihood method to quantify
the significance of the excess in four ranges of the $\epem$ center-of-mass
energy, $\sqrt{s}$, and then
combined the likelihoods.  The net excess has a significance of 
over $6\sigma$, if theoretical uncertainties and correlations for
experimental systematic uncertainties are ignored.  When a conservative
theoretical uncertainty is applied, the significance reduces to $4.4\sigma$,
and when the correlations among experimental measurements are taken
into account, the significance is $4.3\sigma$, corresponding to a
probability of $10^{-5}$.
\par
We compare the data to three straw models of new physics:
1) a purely phenomenological ansatz with $\sigmaNP \propto 1/s$,
2) light bottom squarks ($\sbone$) with little or no coupling to the $Z$, and
3) an additional neutral boson resonance ($\Zprime$) with a mass
well beyond the energies of the data.
We find that the data are consistent with a light $\sbone$, and do
not favor an additional~$\Zprime$.
\par
The structure of this report is as follows.
First, we discuss the data used in this study, and then
the SM prediction for \epemtohad. Next we compare the measurements
to the SM prediction, which is based on the likelihood method.
After establishing the excess on the basis of a very simple
expression for any possible excess, we compare to the expectations
for a light $\sbone$, and also to the expectation for a
heavy $\Zprime$.  Finally, we draw our conclusions.
%
\section{The Data}
\par
We base this analysis on data from the \lepewwg~\cite{lepewwg} and
the compilation of measurements by Zenin \etal~\cite{cscomp}.
\par
The results from the analysis of \lep\ data have been stable for some years.
They fall naturally into two groups: measurements around the $Z$ peak
($88 < \sqrt{s} < 93$~GeV) and well above ($130 < \sqrt{s} < 210$~GeV).
We rely on the reports in Refs.~\cite{lepewwg,lepewwgZ}, 
both for the reduction of the measurements and for the SM predictions.
A summary of the data is given in Table~\ref{lepdata}.
\par
The data below the $Z$ peak come from a number of experiments 
running at several $\epem$ colliders.  The main interest in these
data comes from the need for improved estimates of $\alphaQED(M_Z^2)$
and corrections to \gminustwo.  We consider two subsets, namely,
$20 < \sqrt{s} < 40$~GeV, for which $Z$-exchange should be unimportant,
and $40 < \sqrt{s} < 70$~GeV, which will show the onset of $Z$-exchange.
The combined measurements in these two regions carry approximately equal 
precision.
A summary of the relevant experiments is given in Table~\ref{belowZ}.
\par
\begin{figure}
\begin{center}
\includegraphics[width=0.65\textwidth]{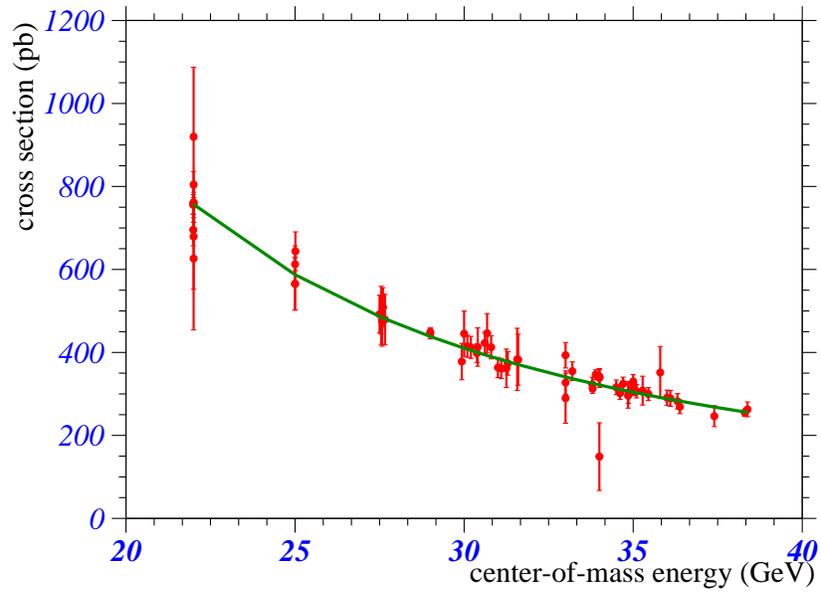}
\caption[.]{\label{region1}
\em Data for region~1 ($20~\GeV < \sqrt{s} < 40~\GeV$) and the
Standard Model prediction from \zfitter.}
\end{center}
\end{figure}
\begin{figure}
\begin{center}
\includegraphics[width=0.65\textwidth]{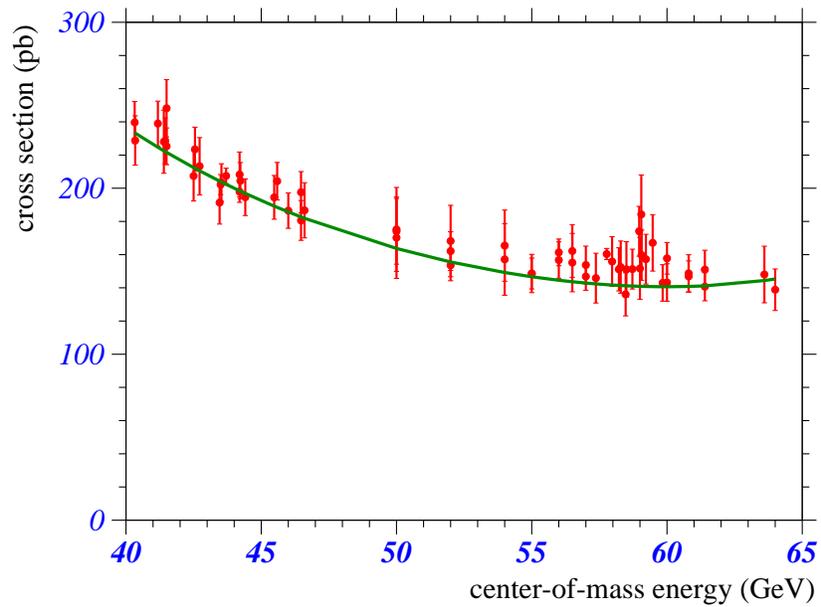}
\caption[.]{\label{region2}
\em Data for region~2 ($40~\GeV < \sqrt{s} < 75~\GeV$) and the
Standard Model prediction from \zfitter.}
\end{center}
\end{figure}
\begin{figure}
\begin{center}
\includegraphics[width=0.65\textwidth]{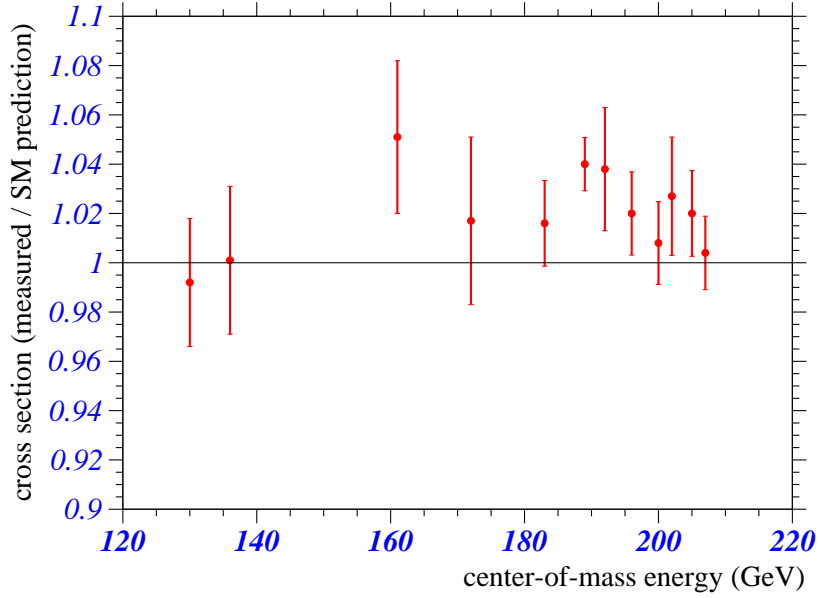}
\caption[.]{\label{region4}
\em Ratio of the \leptwo\ measurements to the SM prediction from
the \lepewwg~\cite{lepewwg}.}
\end{center}
\end{figure}

These considerations lead us to designate four regions based on $\sqrt{s}$:
\begin{center}
\begin{tabular}{cc}
 1 & $20 < \sqrt{s} < 40$~GeV \cr
 2 & $40 < \sqrt{s} < 70$~GeV \cr
 3 & $88 < \sqrt{s} < 93$~GeV \cr
 4 & $130 < \sqrt{s} < 210$~GeV \cr
\end{tabular}
\end{center}
We do not consider the data at $\sqrt{s} < 20$~GeV as
some of these measurements are not very precise, the theoretical uncertainty
from $\alpha_S$ becomes large, and the prediction for any new
particle might be complicated due to threshold effects.
Plots of the data from regions~1, 2~\&~4 are displayed in 
Figs.~\ref{region1},~\ref{region2} and~\ref{region4}.


\par
There are 130 measurements for regions 1 \& 2.  The statistical
precision of the typical measurement is $\sim5\%$, and the total systematic 
uncertainties are roughly the same as the statistical errors.
In some cases the systematic uncertainties are not well described.
Generally speaking, they involve the luminosity measurements, the
hadronization model, and the trigger efficiency.

\begin{table}
\begin{center}
\begin{tabular}{|r|cccc|}
\hline
energy & combined measurement & SM prediction & difference & deviation \cr
\hline
 91.187 &  $41540 \pm 37$ & $41478$ & $62$ & $1.68$ \cr
\hline
 130 & $82.1\pm2.2$ & $82.8$ &     $-0.7$  &  $-0.32$ \cr
 136 & $66.7\pm2.0$ & $66.6$ &     $~0.1$  &  $~0.05$ \cr
 161 & $37.0\pm1.1$ & $35.2$ &     $~1.8$  &  $~1.64$ \cr
 172 & $29.23\pm0.99$ & $28.74$ &  $0.49$  &  $~0.49$ \cr
 183 & $24.59\pm0.42$ & $24.20$ &  $0.39$  &  $~0.92$ \cr
 189 & $22.47\pm0.24$ & $22.16$ &  $0.31$  &  $~1.29$ \cr
 192 & $22.05\pm0.53$ & $21.24$ &  $0.81$  &  $~1.53$ \cr
 196 & $20.53\pm0.34$ & $20.13$ &  $0.40$  &  $~1.18$ \cr
 200 & $19.25\pm0.32$ & $19.09$ &  $0.16$  &  $~0.50$ \cr
 202 & $19.07\pm0.44$ & $18.57$ &  $0.50$  &  $~1.13$ \cr
 205 & $18.17\pm0.31$ & $17.81$ &  $0.36$  &  $~1.16$ \cr
 207 & $17.49\pm0.26$ & $17.42$ &  $0.07$  &  $~0.27$ \cr
\hline
\end{tabular}
\caption[.]{\label{lepdata}
\em Summary of the \lep\ measurements, and SM predictions,
from Ref.~\cite{lepewwg}.
The combined measurements and SM predictions are given in pb.
The `difference' is (measurement - prediction), and the
`deviation' is the difference divided by the experimental error.}
\end{center}
\end{table}

\begin{table}
\begin{center}
\begin{turn}{90}
\begin{tabular}{|lll|}
\hline
experiment & $\sqrt{s}$ (GeV) & publication \cr
\hline\hline
CELLO / PETRA &
   34.9, 42.7
  &
   Phys.Lett {\bf 144B} (1984) 297
  \cr
 &
   22.0, 33.8, 38.3, 41.5, 43.5, 44.2, 46.0, 46.6
  &
   Phys.Lett {\bf 183B} (1987) 400
  \cr
\hline
JADE / PETRA &
   22.0, 25.0, 27.7, 29.9, 30.4, 31.3, 33.9, 34.5,
  &
  \cr
 &
   \hspace*{10pt}  35.0, 35.4, 36.4, 40.3, 41.2, 42.6, 43.5, 44.4, 45.6, 46.5
  &
   Phys.Rep. {\bf 148} (1987) 67
  \cr
 &
   22.0, 27.6, 30.8
  &
   Phys.Rep. {\bf 83} (1981) 151
  \cr
\hline
Mark-J / PETRA &
   31.6
  &
   Phys.Lett. {\bf 85B} (1979) 463
  \cr
 &
   22.0, 27.6, 30.0, 30.7, 31.6, 33.0, 34.0, 35.3, 35.8
  &
   Phys.Rep. {\bf 63} (1980) 337
  \cr
 &
   34.8
  &
   Phys.Lett. {\bf 108B} (1982) 63
  \cr
 &
   22.0, 25.0, 30.6, 33.8, 34.6, 35.1, 36.3, 37.4, 38.4, 
   &
  \cr
 &
   \hspace*{10pt} 40.3, 41.5, 42.5, 43.5, 44.2, 45.5, 47.5
   &
   Phys.Rev. {\bf D34} (1986) 681
  \cr
\hline
TASSO / PETRA &
   22.0, 27.7, 30.9, 31.6
  &
   Z.Phys. {\bf C4} (1979) 87
  \cr
     &
   22.0, 25.0, 27.6, 30.2, 31.0, 33.0, 34.0, 35.0, 36.0
  &
   Phys.Lett. {\bf 113B} (1982) 499
  \cr
     &
   22.0, 25.0, 27.7, 30.1, 31.5, 33.5, 34.5, 35.5, 36.7, 43.1
  &
   Z.Phys. {\bf C22} (1984) 307
  \cr
     &
   41.4, 44.2
  &
   Phys.Lett. {\bf 138B} (1984) 441
  \cr
     &
   22.0, 35.0, 43.7
  &
   Z.Phys. {\bf C47} (1990) 187
  \cr
\hline
Mark-II / PEP &
   29.0
  & 
   Phys.Rev. {\bf D43} (1990) 34
  \cr
\hline
MAC / PEP &
   29.0
  & 
   Phys.Rev. {\bf D31} (1985) 1537
  \cr
\hline
AMY / TRISTAN &
   50.0, 52.0, 54.0, 55.0, 56.0, 56.5, 57.0, 58.5, 
  &
  \cr
 &
   \hspace*{10pt} 59.0, 59.1, 60.0, 60.8, 61.4
  &
   Phys.Rev. {\bf D42} (1990) 1339
  \cr
\hline
TOPAZ / TRISTAN &
   50.0, 52.0, 54.0, 55.0, 56.0, 56.5, 57.0, 58.3, 
  &
  \cr
 &
   \hspace*{10pt} 59.1, 60.0, 60.8, 61.4
  &
   Phys.Lett. {\bf 234B} (1990) 525
  \cr
                &
   57.4, 58.0, 58.2, 58.5, 58.7, 59.0, 59.2, 59.5, 59.8
  &
   Phys.Lett. {\bf B304} (1993) 373
  \cr
                &
   57.8
  &
   Phys.Lett. {\bf B347} (1995) 171
  \cr
\hline
VENUS / TRISTAN &
   50.0, 52.0
  &
   Phys.Lett. {\bf 198B} (1987) 570
  \cr
                &
   63.6, 64.0
  &
   Phys.Lett. {\bf B246} (1990) 297
  \cr
\hline
\end{tabular}
\end{turn}
\caption[.]{\label{belowZ}
\em Relevant data from below the $Z$ peak.
(Energies below 20~GeV are not listed.)}
\end{center}
\end{table}

%
\section{Standard Model Predictions}
\par
\par
The SM predictions for the inclusive process \epemtohad\ are relatively
simple.  Let $M_f$ be the mass of a particular quark, so
$\beta \equiv \sqrt{1 - 4\,M_f^2/s}$ is its velocity.
Denote its electric charge by $Q_f$, and weak isospin by 
$I^f_3 = \pm \frac{1}{2}$.  The fundamental constants are
the Fermi constant, $G_F$, the QED coupling $\alphaQED$,
and the strong coupling $\alpha_S$.  The latter two run
as a function of $s$.  Finally, the weak mixing angle
relates weak and electric couplings, and can be defined 
according to
\begin{equation}
\label{sinsqthetaW}
   G_F\, M_Z^2 = \frac{\pi\alphaQED}
                 {\sqrt{2}\,\ssqW\,\csqW}
\end{equation}
where $M_Z$ is the mass of the $Z$ boson.
Since $\alphaQED(s)$ runs, so does $\ssqW$, according to this definition.
The vector and axial vector couplings of the quark $f$ to the $Z$
can be written
\begin{equation}
\label{va}
 v_f = \frac{I^f_3 - 2 Q_f \ssqW}{2 \ssqW \csqW} \qquad
 a_f = \frac{I^f_3}{2 \ssqW \csqW}
\end{equation}
and the $Z$ propagator in the lowest order is simply~\cite{lep1yb}
\begin{equation}
 \chi_0(s) = \frac{s}{s - M_Z^2 + i\,M_Z\,\Gamma_Z}
\end{equation}
where $\Gamma_Z$ is the $Z$ width.
\par
The amplitude for $\epem \rightarrow \qqbar$  is the sum of the
amplitudes for $\gamma^\star$ and $Z$ exchange, so the cross section
is the sum of three terms.  After integrating over all angles,
\begin{eqnarray}
\label{cs}
 \sigma(\epem \rightarrow \qqbar) &=&
  \frac{2\,\pi\,\alphaQED^2}{3 s}  \cdot   
   \beta \left(1 + \frac{1}{2\gamma^2}\right) \cdot
    C_S \cdot
  \cr & &
  \left[
    Q^2_f  -2 v_e v_f Q_f \, Re \chi_0(s) +
     (v_e^2+a_e^2) 
   \left( v_f^2 + a_f^2 \left( \frac{2+2\gamma^2}{1+2\gamma^2} \right) \right)
      \, \left| \chi_0(s) \right|^2
  \right] \qquad
\end{eqnarray}
where $C_S$ is a QCD correction factor depending on $\alpha_S$ and
the number of colors, $N_C = 3$.
\par
Equation~(\ref{cs}) pertains to the simple Born approximation when
$\alphaQED$, and hence $\ssqW$, are fixed, and $C_S = N_C$.
The {\em improved} Born approximation allows for running $\alphaQED$
and $\ssqW$, for finite corrections to $C_S/N_C$, and for an 
energy dependent width:  $\Gamma_Z = \Gamma^0_Z\, s /M_Z^2$, 
with $\Gamma^0_Z$ independent of~$s$.
We list the values for all required constants in Table~\ref{inputs}.

\begin{table}
\begin{center}
\begin{tabular}{|ll|}
\hline
parameter & value \cr
\hline
 $1 / \alphaQED(0)$     &  $137.036$ \cr
 $\LambdaQCD$           &  $213^{+76}_{-70}$~MeV \cr
 $M_Z$                  &  $91.1876 \pm 0.0021$~GeV \cr
 $\Gamma^0_Z$           &  $2.4952 \pm 0.0023$~GeV \cr
 $M_b$                  &  $4.5$~GeV \cr
 $M_c$                  &  $1.6$~GeV \cr
\hline
\end{tabular}
\caption[.]{\label{inputs}
\em Constants required for the evaluation of the cross section
in the improved Born approximation. The given value for $\LambdaQCD$
leads to $\alpha_S(M_Z^2) = 0.1184 \pm 0.0072$.  The uncertainties
on $\alphaQED$, $M_b$ and $M_c$ are irrelevant for this analysis.}
\end{center}
\end{table}

We use version 6.36 of the \zfitter\ program~\cite{zfitter} to compute 
the SM predictions in  regions 1~\&~2.  This program has been developed 
by several authors over many years, and is one of the two main programs 
employed by the \lepewwg.  For an in-depth discussion of the theoretical
accuracy of the predictions from \zfitter, see the paper by Bardin,
Gr\"unewald, and Passarino~\cite{theoaccy}.
\par
The routine for computing the $s$-dependent hadronic corrections 
to $\alphaQED$ was recently improved by Jegerlehner~\cite{jegerlehner},
who provided us with nearly-final version.  We use this new routine
for our SM computations.  The change in $\dahad$ is generally less
than $2\%$, and so the impact on the cross section is negligible. 
\par
We wrote our own code to compute the cross section
for \epemtohad, based on Eq.~({\ref{cs}).  For $\alphaQED(s)$
we used a routine from H.Burkhardt~\cite{burkhardt}, and for the
QCD correction factor, $C_S$, we used an expression published by
S.Bethke~\cite{bethke}:
\begin{equation}
\label{qcdcorr}
 C_S = N_C \Bigl( 1 + \kappa_1\,\left(\frac{\alpha_S}{\pi}\right)
              + \kappa_2\,\left(\frac{\alpha_S}{\pi}\right)^2 \Bigr)
\end{equation}
with $\kappa_1 = 1$ and $\kappa_2 = 1.4$.
The exact values of $\kappa_1$ and $\kappa_2$ depend slightly on
the quark masses, and we consider the uncertainties of these coefficients below.
\par
We compared the results of our code to those of \zfitter.
There is agreement at the level of $0.2~\%$ after accounting for the
fact that the running of the heavy quark masses is implemented in \zfitter\
but not in our code;  the predictions from \zfitter\  are slightly {\em higher} 
than those from our code at the level of $0.1$--$0.2~\%$.  Also, box diagrams
contribute about $0.5~\%$ to \epemtohad\ above the $W^+W^-$ threshold,
which is not included in the simple expression Eq.~(\ref{cs}).
For the comparison of the SM prediction to the data, we use the
values from \zfitter, as this program is extremely well tested and reliable. 
Our code is useful for investigating sources
of theoretical errors, as discussed below.  
\par
We must assess the uncertainty on the SM prediction in order to
make the comparison to the data.
The studies by Bardin \etal~\cite{theoaccy} indicate that the theoretical 
accuracy is 0.2\% or better, in regions 1~\&~2.
For region 4, the Two-Fermion Working Group reports a theoretical
uncertainty of 0.26\%~\cite{twofermionwg}.  They also report values for
$1/\alphaQED$ which imply an uncertainty on \csepemtohad\ of 0.14\%,
and we obtain similar results.
\par
The uncertainty coming from the inputs must also be estimated.
For $\alphaQED(s)$, we must consider both the uncertainty on its
measured value at $\sqrt{s} \approx 0$, and the uncertainty from 
the running, which is dominated by the uncertainty on $\dahad$.
The uncertainty from $\alphaQED(0)$ is negligible.  
We estimate the uncertainty from the running by modifying the 
coefficients in the Burkhardt routine to correspond to 
$\dahad = 0.02761 \pm 0.00036$  at the $Z$~peak~\cite{burkhardt}.
This is a conservative estimate and is the same used by the
\lepewwg\ in their analyses of \lep\ data.  Other estimates
suggest a significantly smaller uncertainty~\cite{alphaother}.
Furthermore, we take the uncertainty from $\alphaQED$ on the
SM cross section to be fully 
$\delta(\sigma)/\sigma \sim 2\,\delta(\alphaQED)/\alphaQED$,
ignoring the fact that uncertainties will be smaller for the
terms involving $G_F$, as we have explicitly verified using our
code based on Eq.~(\ref{cs}).
\par
The uncertainty coming from the QCD correction also has two pieces:
the uncertainty on $\alpha_S(s)$ and the uncertainty on the coefficient
$\kappa_2$.   We do not consider an uncertainty on $\kappa_1$ as
this would be redundant with the uncertainty on $\alpha_S$.
As discussed by S.Bethke~\cite{bethke}, there is some spread in the
values for $\alpha_S$ coming from various sets of measurements.
There is a tendency for the lower energy measurements, such as
deep-inelastic scattering, to give a value of $\alpha_S(M_Z^2)$
lower than that coming from the data at higher energy, mainly from \lep.  
A typical average is $\alpha_S(M_Z^2) = 0.1184 \pm 0.0036$~\cite{bethke}.  
The central value is more or less mid-way between the low values 
favored by DIS measurements and the high values favored by \lep.  
In the context of the light sbottom model discussed below, 
$\alpha_S$ will run more slowly resulting in a slightly higher 
value at $s = M_Z^2$~\cite{rosner}. We choose to keep the central
value ($0.1184$), but double the uncertainty to $\pm 0.0072$.  In the 
parametrization of the running of $\alpha_S$ by Bethke~\cite{bethke}, 
this corresponds to the low-energy parameter $\LambdaQCD = 213^{+76}_{-70}$~MeV.
We evaluated the impact of this uncertainty on the SM cross section
using our code and find it to be on the order of $\delta(\sigma)/\sigma
\sim 0.3\%$.  We also varied the coefficient
$\kappa_2$ by 10\% and found the impact to be negligible.
\par
We evaluated the uncertainties coming from the other input parameters,
namely, $M_Z$, $\Gamma^0_Z$, $M_b$ and $M_c$, and found them
to be completely negligible.
\par
A summary of the theoretical uncertainties for the entire range of $\sqrt{s}$
is given in Fig.~\ref{plottheounc}.  The dominant error comes
from the running of $\alpha_S$, which has been very conservatively
estimated.  We display also the {\em linear sum} of the uncertainties,
and use this as the guide for setting the theoretical uncertainty.
For the range $20~\GeV < \sqrt{s} < 75~\GeV$,  we take an uncertainty 
on the order of 0.5\%, which falls slightly from $20~\GeV$ to $80~\GeV$,
as indicated by the heavy line.
For region~3, we do not include a theoretical uncertainty,
as this parameter is a fit result, and the reported theoretical
uncertainty is exceedingly small, amounting to $0.012\%$, or $5~\pb$~\cite{theoaccy}.
In region~4, we take an uncertainty of $0.55\%$, as indicated by
the second heavy line.
\par
We emphasize that these theoretical uncertainties are conservative, and one
could argue that more agressive estimates closer to $0.3~\%$ would be
justified.

\begin{figure}
\begin{center}
\includegraphics[width=0.8\textwidth]{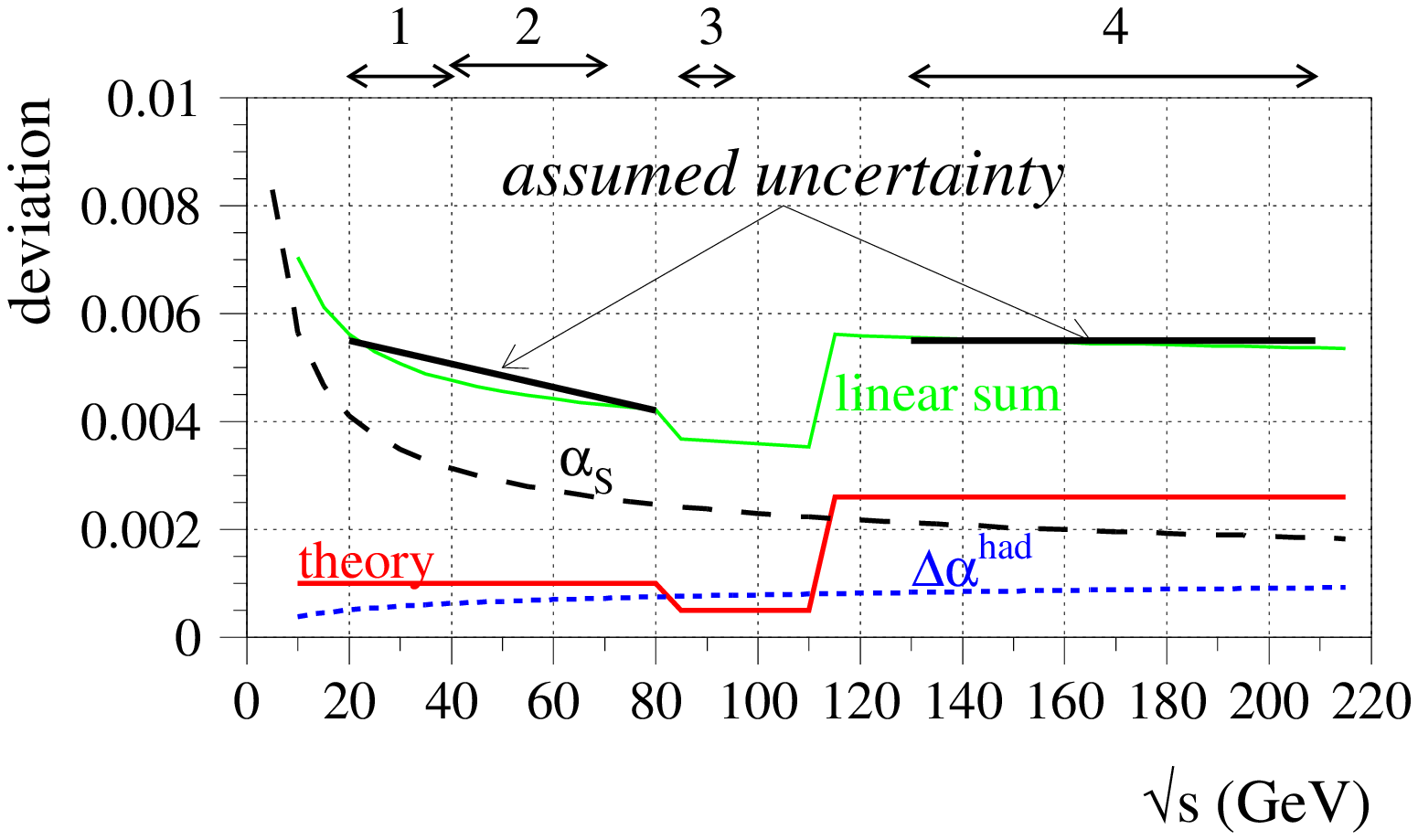}
\caption[.]{\label{plottheounc}
\em Theoretical uncertainties in the Standard Model prediction
for \csepemtohad.}
\end{center}
\end{figure}

%
\section{Comparison of the Data to the SM Prediction}
\par
We develop a series of comparisons, starting with a simple weighted
average of differences and ending with a likelihood analysis which
incorporates theoretical and experimental systematic uncertainties,
and correlations among errors.
\subsection{Mean Deviations}
\par
We compare the SM prediction to the data in each of the four 
kinematic regions defined above.
First, we compute a simple $\chi^2$ and mean deviation of the
data from the SM prediction.  Let the measurements be $y_i$
with uncertainties $\eta_i$, and the SM prediction, $\ySM$,
which varies with~$s$.
Then
\begin{equation}
\label{chisq}
 \chi^2 = \sum_{i=1}^{N} \left( \frac{y_i-\ySM}{\eta_i} \right)^2
\end{equation}
and the mean deviation and rms are
\begin{equation}
\label{meandev}
   \bar{\Delta} \equiv
    \sum_{i=1}^{N} \left( \frac{y_i-\ySM}{\eta_i^2} \right)  
   \Bigl/
    \sum_{i=1}^{N} \left( \frac{1}{\eta_i^2} \right)  
   \qquad {\mathrm{and}} \qquad
   \sigma_{\bar{\Delta}} \equiv
    \left[ \sum_{i=1}^{N} \left( \frac{1}{\eta_i^2} \right) \right]^{-1/2}  .
\end{equation}
At this point we ignore correlations among the measurements.
We summarize the values for $\chi^2$, $\bar{\Delta}$ and $\sigma_{\bar{\Delta}}$
in Table~\ref{meandevvalues}.  No value is given for region~3 as there
is only one data point there.
Note that the mean deviation is significantly non-zero and positive,
even though the $\chi^2$ is fine.  This is a key point and a somewhat
atypical situation.  There are many measurements, each of which is
statistically consistent with the SM prediction, and so the 
value of $\chi^2$ is good.  But the $\chi^2$ test does not distinguish
between measurements which are higher than the prediction and those
which are lower.  In the present case, it turns out that there are
many more which are higher than lower.  The mean deviation, $\bar{\Delta}$,
is designed to make this evident.  
To demonstrate that these offsets are plausible, we form local averages
of the data for regions 1~\&~2, and plot the ratio of the rebinned data
to the SM prediction -- see Fig.~\ref{rebinned}.
The likelihood analysis discussed in the 
next section provides a more physical basis for analyzing the apparent
excess indicated by the values in Table~\ref{meandevvalues}.

\begin{table}
\begin{center}
\begin{tabular}{|c|cc|ccc|}
\hline
region & N data points & $\chi^2$ & 
   $\bar{\Delta}$ & $\sigma_{\bar{\Delta}}$ &
   $\bar{\Delta} / \sigma_{\bar{\Delta}}$ \cr
\hline
1 & 67 & 50.4  &   6.7 &  2.5 & 2.9 \cr
2 & 63 & 55.1  &   7.9 &  1.4 & 5.9 \cr
4 & 12 & 12.2  &  0.32 &  0.11 & 2.8 \cr
\hline
\end{tabular}
\caption[,]{\label{meandevvalues}
\em Values for $\chi^2$, $\bar{\Delta}$ and $\sigma_{\bar{\Delta}}$
for regions 1, 2~\&~4. $\bar{\Delta}$ and $\sigma_{\bar{\Delta}}$
are quoted in~pb.}
\end{center}
\end{table}

\begin{figure}
\begin{center}
\includegraphics[width=0.8\textwidth]{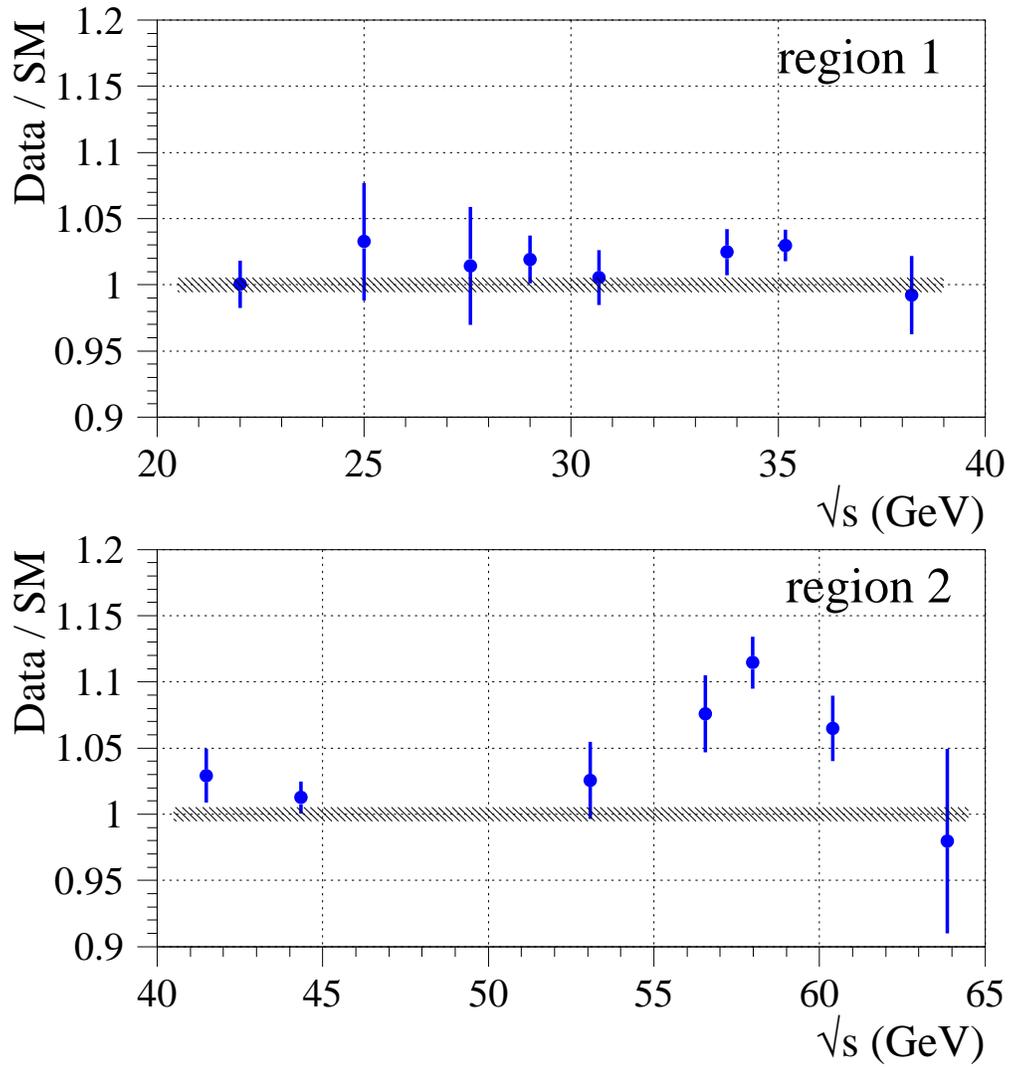}
\caption[.]{\label{rebinned}
\em Ratio of the measurements to the SM prediction, rebinned,
for regions 1~\&~2.  The shaded band centered on~1 represents the
theoretical uncertainty.}
\end{center}
\end{figure}

\subsection{Simple Likelihood Analysis}
\par
We wish to quantify better any disagreement of the SM prediction
with the data.
To this end, we define a likelihood function, based on the
measurements $y_i$ with uncertainties $\eta_i$ and the
theoretical prediction, $\ySM + \alpha\yNP$ (SM + new physics),
where $\alpha$ is a free parameter,
\begin{equation}
\label{likedef}
  {{\cal{P}}} \equiv \prod_{I=1}^N
  \frac{1}{ \sqrt{2\pi}\, \eta_i } \,
 \exp\left( -\frac{1}{2\eta_i^2} (y_i-(\ySM+\alpha\yNP))^2 \right)  .
\end{equation}
It is understood that both $\ySM$ and $\yNP$ are functions of~$s$.
It is more convenient to work with the negative
log-likelihood function, ${\cal{F}} \equiv -\ln {\cal{P}}$.
We will examine the change in ${\cal{F}}$ as a function of $\alpha$.
\par
The quantity $\alpha\,\yNP$ represents a contribution from new physics.
As the simplest ansatz, we take
\begin{equation}
\label{oneovers}
     \yNP(s) \equiv (10~{\mathrm{pb}}) \times
                    \frac{(30~\GeV)^2}{s}
\end{equation}
which has the generic $1/s$ dependence of the pair production
of light particles, and we normalize to 10~pb at $\sqrt{s} = 30~\GeV$
for convenience later.  In this case the influence of $Z$-exchange is
ignored.  Minimizing the negative log-likelihood function
${\cal{F}}$ as a function of $\alpha$, we obtain the results ($\alphabest$) 
summarized in Table~\ref{Fvalues}.  One sees that a positive contribution $\alpha > 0$
is favored in all cases -- this comes about because the mean deviations,
$\bar{\Delta}$ are all positive (see Table~\ref{meandevvalues}).
One also sees that ignoring $Z$-exchange does not appear to be justified,
since $\alphabest$ is much larger for region~3 than for the others.
\par
We compare the best values of ${\cal{F}}$ to those obtained for the SM alone
(\ie, with $\alpha = 0$) and see an improvement when a nonzero ``new physics''
contribution is included.  This improvement can be described by a number
of standard deviations (S.D.) according to
\begin{equation}
\label{SDdef}
   {\mathrm{SD}} =  \sqrt{ 2\,\Bigl[ {\cal{F}}(0) - {\cal{F}}(\alphabest) \Bigr] }
\end{equation}
as discussed in the RPP, for example~\cite{rpp}.
We checked that the minimization of ${\cal{F}}$ and the interpretation
of the error on $\alpha$ corresponded exactly to what is obtained
with a standard $\chi^2$ minimization, region by region.

\begin{table}
\begin{center}
\begin{tabular}{|c|cccc|}
\hline
region & $\alphabest$ & ${\cal{F}}(\alphabest)$ & ${\cal{F}}(0)$ & S.D. \cr
\hline
1 & 0.74 &  21.89 &  25.23 & 2.58 \cr
2 & 1.80 &  25.52 &  37.50 & 4.90 \cr
3 & 57.3 &   0    &  1.404 & 1.68 \cr
4 & 1.37 &  2.00 &  6.12   & 2.87 \cr
\hline
net & 1.183 & 53.446 & 70.255 & 5.80 \cr
\hline
\end{tabular}
\caption[,]{\label{Fvalues}
\em Results of the minimization of the negative log-likelihood function,
${\cal{F}}$, as a function of $\alpha$, for the simple ansatz in
Eq.~(\ref{oneovers}).}
\end{center}
\end{table}

\par
The values listed in Table~\ref{Fvalues} do suggest that there is an
apparent excess of data over the SM prediction, in all four regions 
of~$\sqrt{s}$.  It makes sense to form a combined likelihood by
taking the product of the likelihoods for the four regions.
The corresponding negative log-likelihood function is then
minimized with respect to the scale $\alpha$, with the result:
\begin{equation}
\label{netalpha}
 \alphanet =  1.18 \pm 0.20
 \qquad {\mathrm{or,}} \qquad
     \yNP(s) \equiv (11.8\pm 2.0)~{\mathrm{pb}} \times
                    \frac{(30~\GeV)^2}{s} .
\end{equation}
The difference in ${\cal{F}}$ is 16.81, which corresponds
to 5.8~standard deviations, according to Eq.~(\ref{SDdef}).

\par
The value Eq.~(\ref{netalpha}) is significantly  different from zero,
but still is very small.  It corresponds to only 3\% of the hadronic
cross section at $\sqrt{s} = 30~\GeV$.   At this energy, the cross section
for the production of a lepton pair is about $110~\pb$, and for a
charge $-1/3$ quark, about $40~\pb$.   Interestingly, the cross section for 
a scalar particle will be significantly smaller, due to the reduced
number of spin degrees of freedom.

\subsection{Theoretical Uncertainties and Experimental Systematics}
\par
The theoretical uncertainties depicted in Fig.~\ref{plottheounc} are
not negligible compared to the deviations in Table~\ref{meandevvalues}.
Furthermore, we have taken the measurements to be uncorrelated, which
is not the case.  We now rectify these deficiencies and see to what
extent the significance of the apparent excess diminishes.
\par
To implement the theoretical uncertainties, we modify the likelihood function
to include a multiplier, $\rho$, for $\ySM$ which is constrained by a Gaussian centered
on one with a width given by the assumed theoretical uncertainty, $\eta_\rho$:
\begin{equation}
\label{likedef2}
  {{\cal{P}}} \equiv \left[ 
   \prod_{I=1}^N
 \exp\left( -\frac{1}{2\eta_i^2} (y_i-(\rho\ySM+\alpha\yNP))^2 \right)
   \right]
  \times
 \exp\left( -\frac{1}{2\eta_\rho^2} (\rho-1)^2 \right)
\end{equation}
where we have omitted unnecessary normalization factors.
The likelihood and corresponding negative log-likelihood are to be viewed
as functions of both the new physics scale factor, $\alpha$, and the
scale factor for the SM prediction, $\rho$.  The second factor will allow
a numerical improvement in the likelihood when $\alpha = 0$, but will have
no impact for the optimal $\alphabest$.  The result is a smaller difference
$\Delta{\cal{F}}$ and hence a lower significance.
\par
We apply this new formulation of ${\cal{F}}$ and find a significant
reduction in $\Delta{\cal{F}}$, as expected.  The significance is reduced
from $6\sigma$ to $4\sigma$.
\par
Next we take into account the correlation among the measurements.
For the \leptwo\ data, this is straight forward, as the \lepewwg\
have published the covariance matrix~\cite{lepewwg}.
For the lower energy data, no rigorous prescription is possible,
so we adopt the following method.  The data appear in twenty-one
distinct publications, as listed in Table~\ref{belowZ}.
It is reasonable to assume that the measurements reported in
a given publication are correlated, so we take half the total
systematic uncertainty to be correlated.  Correlations among
sets of measurements, however, are expected to be very small,
since nearly all of the systematic uncertainties come from
detector specific issues (such as the alignment of the
luminosity detectors, or the measurement of the trigger
efficiency), or are evaluated in different ways
(such as the dependence on the hadronization model).
\par
We replace the sum over individual terms in the negative
log-likelihood function by one-half the $\chi^2$ formed
in the canonical way from the inverse of the covariance
matrix.  This $\chi^2$ function depends on $\alpha$
in the same way ${\cal{F}}$ does.
The data in regions 1~\&~2 are treated together
as part of one $\chi^2$ function, and the data
in region~4 is treated in a separate $\chi^2$ function.
\par
Since the weight of each measurement is different once
correlations are taken into account, the net ${\cal{F}}$
function must be minimized anew.  Setting aside the
theoretical uncertainties, the significance of the excess
is reduced from $6\sigma$ to $5\sigma$.  Taking both
the theoretical uncertainties and the correlations into
account, we find a significance of $3.9\sigma$.

%
\section{Examples: Light Sbottoms, Extra $Z$ Boson}
\par
For the sake of discussion, we take the apparent excess to be genuine
and compare the data to two ans\"atze for physics beyond the SM:
\begin{enumerate}
\item {\sl light bottom squarks}
\item {\sl heavy $\Zprime$ bosons}
\end{enumerate}
The first model serves as the example of the production of new light
particle in $\epem$ collisions, while the second involves a new 
production mechanism of purely SM particles.  They can be distinguished,
in the absence of an analysis of the final state, by their dependence
on~$s$: the one falls as $1/s$ while the other rises approximately
as~$s$, far below the pole in the propagator.
We do not mean to advocate either of these models, but rather use them as
reasonable models to test with these measurements of \csepemtohad.
They are both well  motivated by other considerations.

\subsection{Light Sbottoms}
\par
The light sbottom model comes from the proposal by 
E.Berger \etal~\cite{bergeretal} to explain
the excess of $b$-hadron production at the Tevatron by postulating
the existence of a light sbottom quark ($\sbone$), with a mass in 
the range $2~\GeV < \Msbone < 6~\GeV$, and a light gluino ($\gluino$)
with a mass of roughly $\Mgluino \sim 12~\GeV$.
The gluino decays only to $\gluino\rightarrow b\sbone$, and the sbottom 
decays via an $R$-parity violating coupling to a pair of quarks:
$\sbone\rightarrow \bar{u}\bar{d}$, for example.  For our purposes,
the gluino plays no role -- we assume that these light
these bottom squarks produce hadronic jets through an $R$-parity
violating coupling ($\lambda^{\prime\prime}$) which would allow them to be
selected in the measurements of \csepemtohad\  with approximately
the same efficiency as a SM event $\epem\rightarrow\qqbar$.
While the SM $b$ quarks have a substantial coupling to the $Z$ boson,
the composition of the lightest sbottom can be tuned so make this
coupling arbitrarily small.  In the case that it is exactly zero,
the cross section would have the generic $s$-dependence of Eq.~(\ref{oneovers}).
The composition is controlled by a mixing factor, $\cossbone$,
and zero $Z-\sbone-\sbone$ coupling corresponds to $\cossbone = 0.39$.
In our comparison of the light sbottom model to the data, we will
have to allow $\cossbone$ to be a free parameter.  It is known from
a comparison to the total $Z$ width, however, that 
$\cossbone < 0.6$~\cite{argsbot}.
We use the code \msmlib\ from G.Ganis to compute the sbottom
cross sections~\cite{msmlib}, and fix the sbottom mass to $6~\GeV$.
\par
To examine the dependence of ${\cal{F}}$ on $\cossbone$, we first
fix $\alpha \equiv 1$.  A shown below, this is close its optimal value.
The results of the minimization of ${\cal{F}}$ are as one would expect.
The first region is dominated by photon exchange, and so there is no sensitivity
to $\cossbone$.  The second retains a very mild sensitivity to
$\cossbone$, favoring $\cossbone \sim 1$, since the excess in this
region is large.  Region~3 is dominated by $Z$~exchange, and so
the dependence of the cross section on $\cossbone$ is strong.
It clearly favors a reduced coupling to the~$Z$, and the limit
$\cossbone < 0.6$ is easily obtained.  Finally, the data from 
region~4 have a mild sensitivity to $\cossbone$, similar to
what is seen for region~3.

\par
We sum the negative log-likelihood functions for the four regions
and plot the sum as a function of $\cossbone$, as shown in 
Fig.~\ref{fvscossbone}.  This variation is dominated by the
$Z$~pole data.  The preferred value is $\cossbone = 0.18$,
which gives a small but non-zero coupling to the~$Z$, as required
by the small excess in $\cspeak$.  A zero coupling corresponds
to $\cossbone = 0.38$, and is {\em not} excluded by the data --
it is disfavored by only 1.3~standard deviations.

\begin{figure}
\begin{center}
\includegraphics[width=0.8\textwidth]{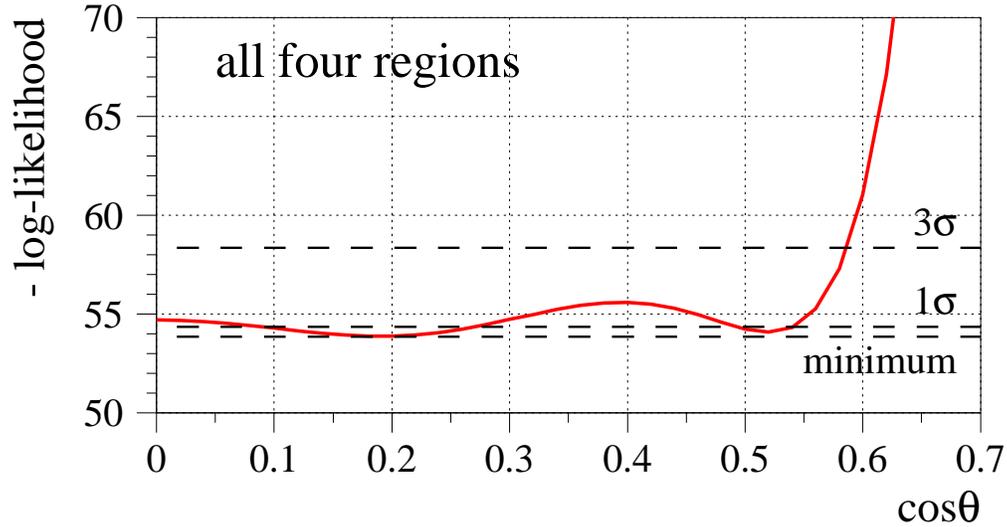}
\caption[.]{\label{fvscossbone}
\em Sum of negative log-likelihood functions as a function of $\cossbone$.
The dashed lines indicate the minimum of ${\cal{F}}$, and the $1\sigma$
and $3\sigma$ levels.}
\end{center}
\end{figure}

\par
We now fix $\cossbone = 0.18$, and return to the procedure to evaluate
the significance of the excess as described in the last section.
Specifically, we allow for an overall theoretical uncertainty, and
we take correlations among measurements into account.
In the present context we are comparing the quality of the description
of the data for SM + a light sbottom to the SM alone.
We vary $\alpha$ freely for the sum of negative log-likelihoods, 
and obtain $\alphabest = 1.17$ for which ${\cal{F}}(\alphabest) = 50.83$,
to be compared to ${\cal{F}}(0) = 60.03$.  In the latter case, the
systematic uncertainty on the SM prediction allows an enhancement
of the prediction by 0.87\% -- \ie, $\rho_{\mathrm{best}} = 1.0087$
in Eq.~(\ref{likedef2}).  This difference in ${\cal{F}}$
corresponds to $4.3$~standard deviations, for a probability of
$9.0 \times 10^{-6}$.
\par
It is important to note the following three points:
\begin{enumerate}
\item
All four regions contribute positively
to $\Delta{\cal{F}}$ at $\alphabest$.
\item
The best overall point, $\alphabest$, is not
disfavored by any of them, and is statistically
consistent with the best point for each individual region.
\item
The best overall point is consistent with~$\alpha = 1$.
\end{enumerate}
To illustrate these points graphically, we temporarily revert
to the definitions of ${\cal{F}}$ which neglect theoretical
uncertainties and correlations, and plot the variation of
${\cal{F}}$ for each of the four regions and their sum --
see Fig.~\ref{fvsalpha}.  
\par
The minima of the four parabolas fall closer to $\alpha = 1$
than to $\alpha = 0$.  This indicates there is a consistent 
excess in all four regions.  The sum of the four parabolas
has a sharper minimum at $\alpha = 1.2$.   The rise in the
negative log-likelihood passing from $\alpha = 1.2$ to 
$\alpha = 0$ corresponds, when interpreted as a number
of standard deviations, to $6\sigma$.  
(Recall that, after taking theoretical uncertainties and
correlations among measurements into account, this number
is reduced to $4.3\sigma$.)
In this sense one can say that the data prefer a description 
which includes a light sbottom, with a high level of
statistical significance.  
If the parabola were centered at $\alpha = 0$ instead,
one would say that this model for a light sbottom is
excluded.  However, one cannot say that the data exclude 
the Standard Model.  The contribution to \epemtohad\
of SM processes is indisputable, while the contribution
from light sbottoms is speculative.  The results of
this analysis do not prove the existence of light sbottoms,
rather, they suggest there may be new processes 
contributing to the hadronic final state.

\begin{figure}
\begin{center}
\includegraphics[width=0.9\textwidth]{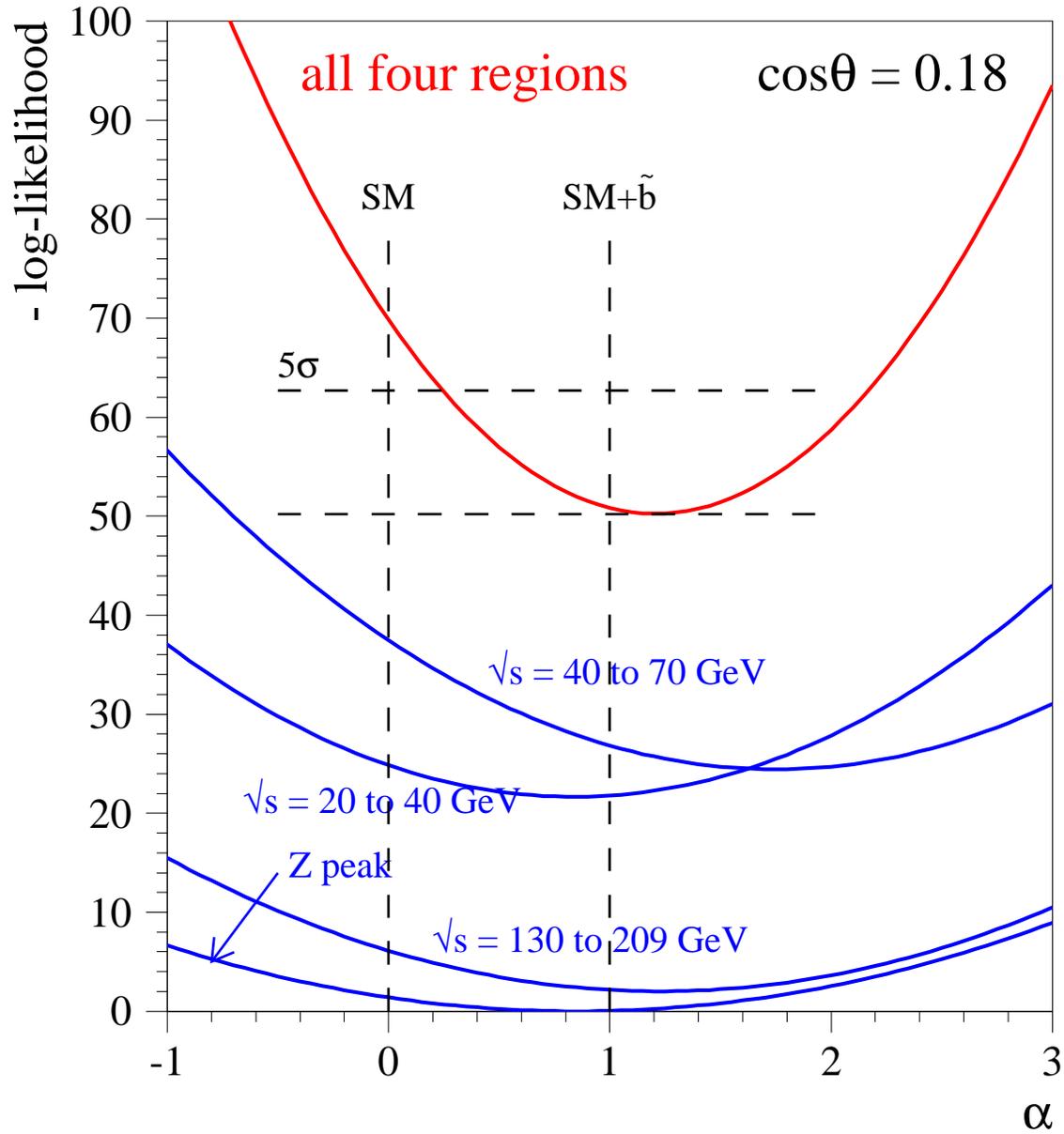}
\caption[.]{\label{fvsalpha}
\em Negative log-likelihood as functions of $\alpha$.  The mixing angle
has been fixed by $\cossbone = 0.18$.  The four lower parabolas show
the contributions from each of the four regions, as marked.  Their sum
results in the uppermost parabola, which has a minimum close to
$\alpha = 1$, corresponding to the sum of SM and $\sbone$ cross
sections.  This point is preferred over the SM alone $\alpha = 0$.}
\end{center}
\end{figure}

For illustration, we plot the difference between the rebinned measured values
and the SM prediction as a function of $\sqrt{s}$, as shown in 
Fig.~\ref{dsigplot}.  The shaded band centered on zero indicates
the conservative error we have assigned to the SM prediction.
The heavy smooth curve indicates the cross section for a light
sbottom ($\Msbone = 6~\GeV$) when $\cossbone = 0.18$.

\begin{figure}
\begin{center}
\includegraphics[width=0.9\textwidth]{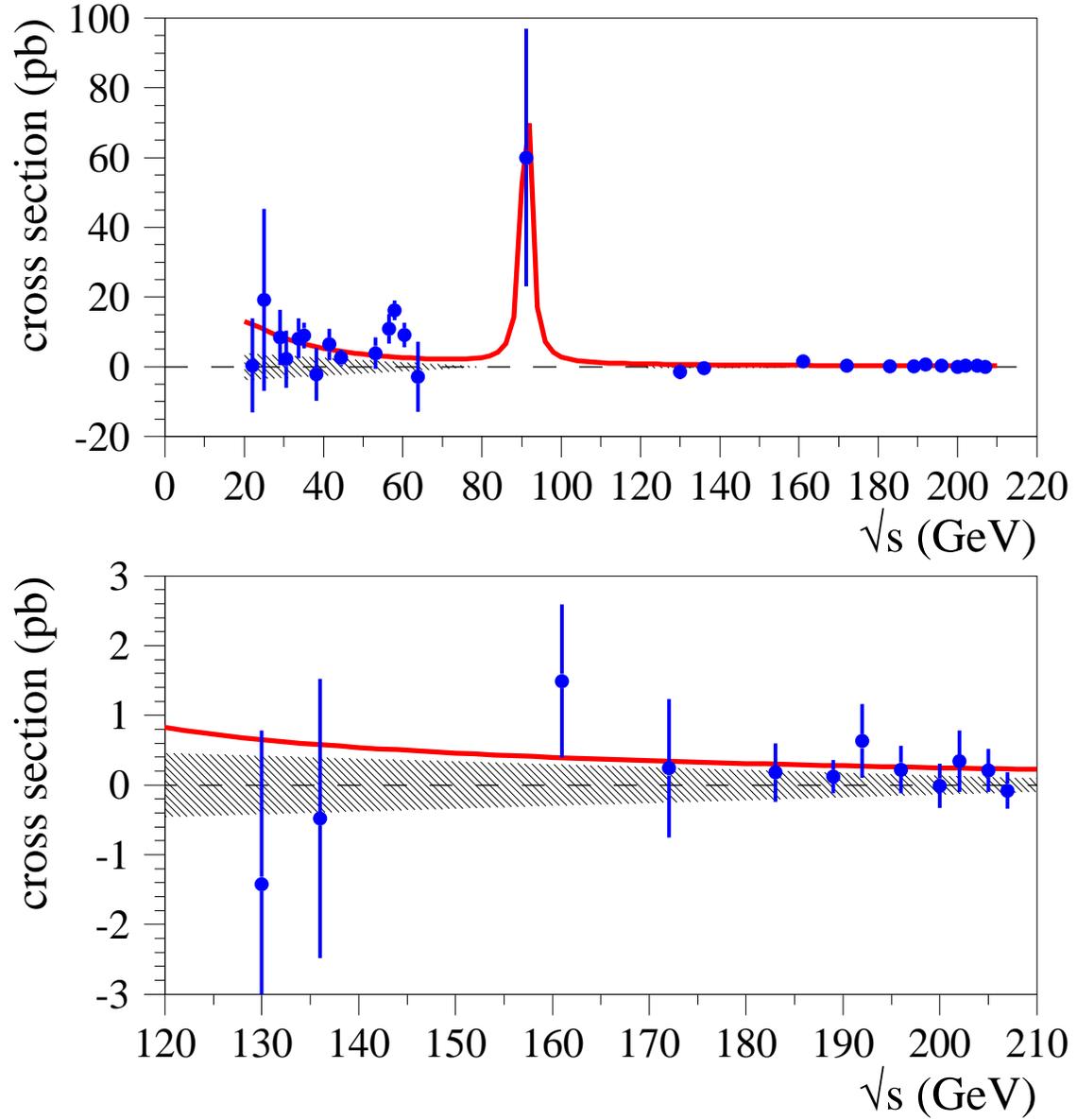}
\caption[.]{\label{dsigplot}
\em Comparison of the data to the prediction of a light sbottom
($\Msbone = 6~\GeV$) with $\cossbone = 0.18$.
The points represent the data, rebinned in the range
$20~\GeV < \sqrt{s} < 75~\GeV$.  The upper plot shows the full
range of $\sqrt{s}$, while the lower plot shows the \leptwo\ data
alone.}
\end{center}
\end{figure}

\newpage

\subsection{A Heavy $\Zprime$ Boson}
\par
We consider the exchange of a third boson, call it $\Zprime$,
in the process \epemtohad.  For simplicity, we assume that
there is no interference between this $\Zprime$ and the photon
and SM $Z$ boson.  If the mass of this $\Zprime$, $\MZprime$, were
much higher than the $\sqrt{s}$ probed by these measurements,
then cross section would rise approximately linearly in $s$:
\begin{equation}
\label{linears}
  \sigma(s) \sim \frac{A}{s} \left|  
       \frac{s}{s - M_{Z^\prime}^2 + i\,M_{Z^\prime}\,\Gamma_{Z^\prime} }
                 \right|^2
 \approx A \, \frac{s}{M_Z^4} 
  \qquad
  A = \frac{2}{3}\pi\alphaQED^2 C_S 
  \left( (v_e^\prime)^2 +(a_e^\prime)^2 \right)
  \left( (v_f^\prime)^2 +(a_f^\prime)^2 \right)  .
\end{equation}
The factor $A$ contains the coupling and other factors.  
\par
We use the expression Eq.~(\ref{likedef}) with
$\yNP(s) = (10~\pb) \cdot (s/(300~\GeV)^2)$.
A summary of the results for the four regions is given in 
Table~\ref{FvaluesZP}.
The combined likelihood gives $\alphanet = 0.00072$, for 
a $\Delta{\cal{F}} = 3.74$, corresponding to 2.8~standard
deviations.  
This model makes only a small improvement over the SM.
The rising $s$ dependence pits the \leptwo\ data against the
data in regions 1~\&~2 -- one has either $\yNP$ too high
in region~4 or too small in regions 1~\&~2.
This model is distinctly less successful than the other two.

\begin{table}
\begin{center}
\begin{tabular}{|c|cccc|}
\hline
region & $\alphabest$ & ${\cal{F}}(\alphabest)$ & ${\cal{F}}(0)$ & S.D. \cr
\hline
1 & 0.50   &  21.89 &  25.23 & 2.58  \cr
2 & 0.27   &  17.31 &  37.50 & 6.35  \cr
3 & 0.67   &   0    &  1.404 & 1.68  \cr
4 & 0.0007 &  2.46  &  6.12   & 2.70 \cr
\hline
\end{tabular}
\caption[,]{\label{FvaluesZP}
\em Results of the minimization of the negative log-likelihood function,
${\cal{F}}$, as a function of $\alpha$, for the case of a heavy $\Zprime$.}
\end{center}
\end{table}

%
\section{Summary}
\par
We have examined the measurements of \csepemtohad\ across a wide
range of energies and experiments.  None of these measurements
shows a significant excess over the predictions of the Standard Model.
Taken together, however, there is a clear trend, which has been
quantified in a  number of ways.
\par
Motivated by this apparent excess, we compared the data to a model
for the production of light scalar $b$-quarks,
of the type suggested by the Argonne group~\cite{bergeretal}.
As a specific example, we have considered sbottoms with a mass of
$6~\GeV$ and a mixing of left- and right-states parametrized by
$\cossbone = 0.18$.  This mixing gives a relatively small
but non-zero coupling to the $Z$~boson.
\par
When all measurements are taken to be independent, and when
theoretical uncertainties in the SM prediction are ignored,
the light sbottom model is `preferred' over the SM by $6.1\sigma$.
When correlations among measurements are taken into account,
this significance drops to $5.2\sigma$.  Finally, when a very
conservative theoretical uncertainty is folded into the calculation,
the resulting significance is $4.3\sigma$, which corresponds to
a probability of $9\times 10^{-6}$.
\par
Alternative models are less successful in describing the data.
A new neutral gauge boson is disfavored, and a cross section 
with a generic $1/s$ dependence does not explain the slight
excess at \lepone\ (though it still provides an improvement
over the SM with a significance of $3.9\sigma$).
\par
We conclude that the description of the data on \epemtohad\ is
significantly improved if Standard Model processes are
augmented by the pair production of light scalar particles of
charge $\pm 1/3$ which appear as hadronic final states.


}  

\vskip 0.5in
\noindent
{\large\bf Acknowledgments}
\vskip 0.125in
\noindent
I am very grateful to Andr\'e de Gouv\^ea for many stimulating discussion
through the course of this study.  I also acknowledge helpful discussions 
with Ed~Berger, Marcela~Carena, Martin~Gr\"unewald, Fred~Jegerlehner,
Zack~Sullivan, Tim~Tait, Mayda~Velasco and Carlos~Wagner.
This work was supported by US DoE contract DE-FG02-91ER40684.

\end{document}